\documentclass{ws-procs9x6}

\begin{document}

\title{Transversity and Transverse-Momentum-Dependent Distribution Measurements from PHENIX and BRAHMS}

\author{C. Aidala$^*$ for the PHENIX and BRAHMS Collaborations}

\address{Physics Department, University of Massachusetts,\\
Amherst, MA 01003, U.S.A.\\
$^*$E-mail: caidala@bnl.gov}

\begin{abstract}
A variety of measurements performed utilizing transversely polarized proton-proton collisions at the Relativistic Heavy Ion Collider (RHIC) are now available.  Recent results from the PHENIX and BRAHMS experiments are presented and discussed.
\end{abstract}

\keywords{transverse spin; proton structure.}

\bodymatter

\section{Introduction}
Surprisingly large transverse single-spin asymmetries (SSA's) were initially observed in hadronic collisions at the ZGS at Argonne National Laboratory \cite{Klem:1976ui,Dragoset:1978gg} as well as the PS at CERN \cite{Antille:1980th} in the late 1970's and early 1980's, for center-of-mass energies between 5 and 10~GeV.  Subsequent experimental data over a wide range of energies indicate that similar effects exist for particle production at center-of-mass energies of tens of GeV and even as high as 200~GeV. The higher-energy data allow perturbative QCD (pQCD) to be used in attempting to interpret the results.  While much remains to be learned, exciting progress has been made in the understanding of the observed transverse single-spin asymmetries at high energies, and the wealth of experimental data now becoming available is helping to drive the field forward.

The complex task of understanding measurements performed in hadronic collisions is now being facilitated as relevant quantities are being studied in semi-inclusive deep-inelastic scattering (SIDIS) and $e^+e^-$ annihilation.  A definitively non-zero Collins fragmentation function (FF) measured in $e^+e^-$ annihilation was recently published by the BELLE experiment \cite{Abe:2005zx}.  The availability of the Collins FF makes it possible to extract the transversity distribution from asymmetry measurements in SIDIS and proton-proton collisions, and first extractions of transversity have been published \cite{Anselmino:2007fs}, utilizing the available SIDIS data on the proton from HERMES \cite{Airapetian:1999tv,Airapetian:2004tw} and the deuteron from COMPASS \cite{Alexakhin:2005iw}.  First extractions of the transverse-momentum-dependent Sivers distribution function from these SIDIS data have also been released \cite{Anselmino:2008sga,Arnold:2008ap}.  As constraints start to be provided on transversity and the various transverse-momentum-dependent distribution and fragmentation functions, more can in turn be learned from hadronic collision data.

\section{Using pQCD to describe polarization-averaged cross sections at RHIC}

\begin{figure}
\begin{center}
\includegraphics[width=0.45\textwidth,height=0.3\textheight]{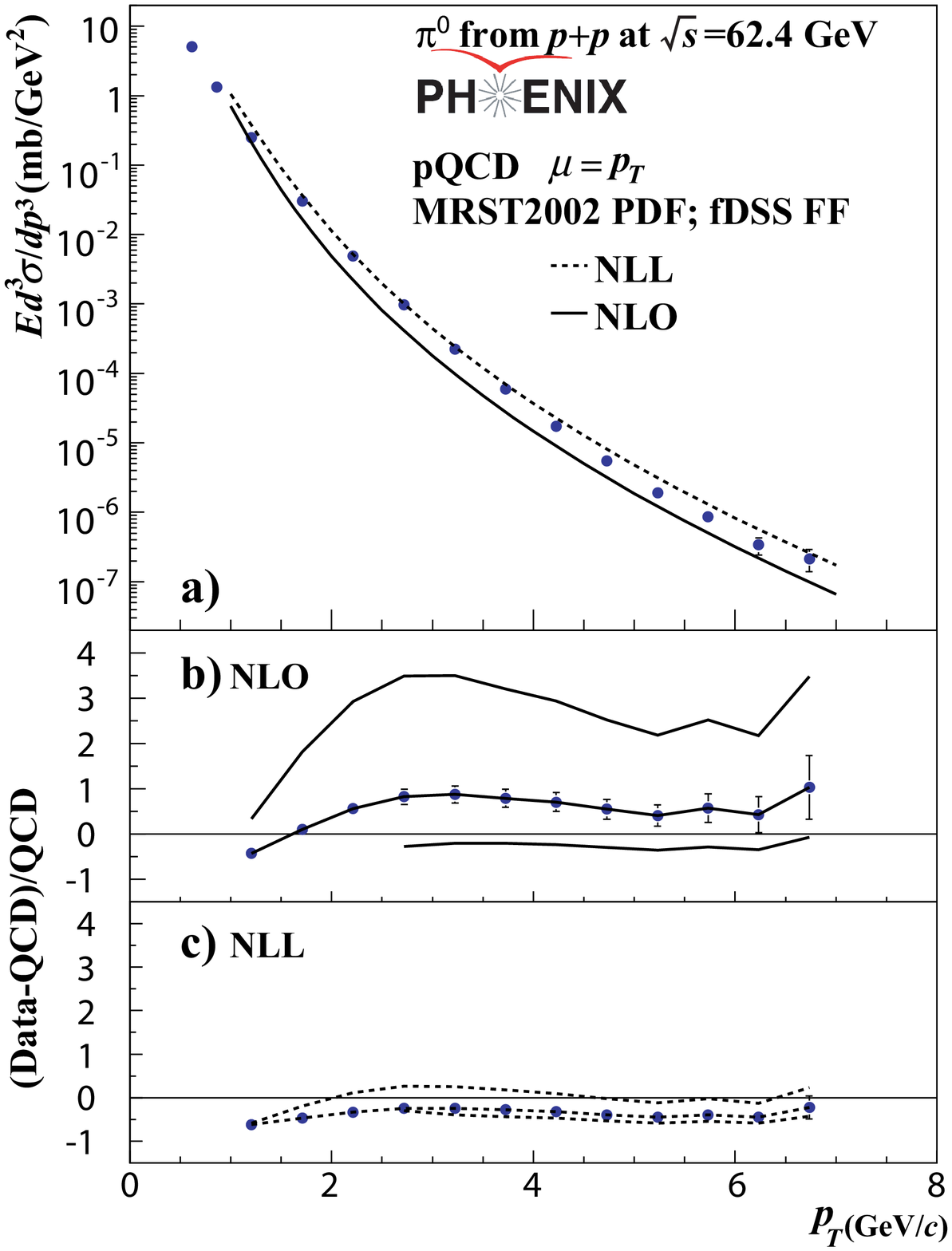}
\hspace{0.03\textwidth}
\includegraphics[width=0.47\textwidth,height=0.33\textheight]{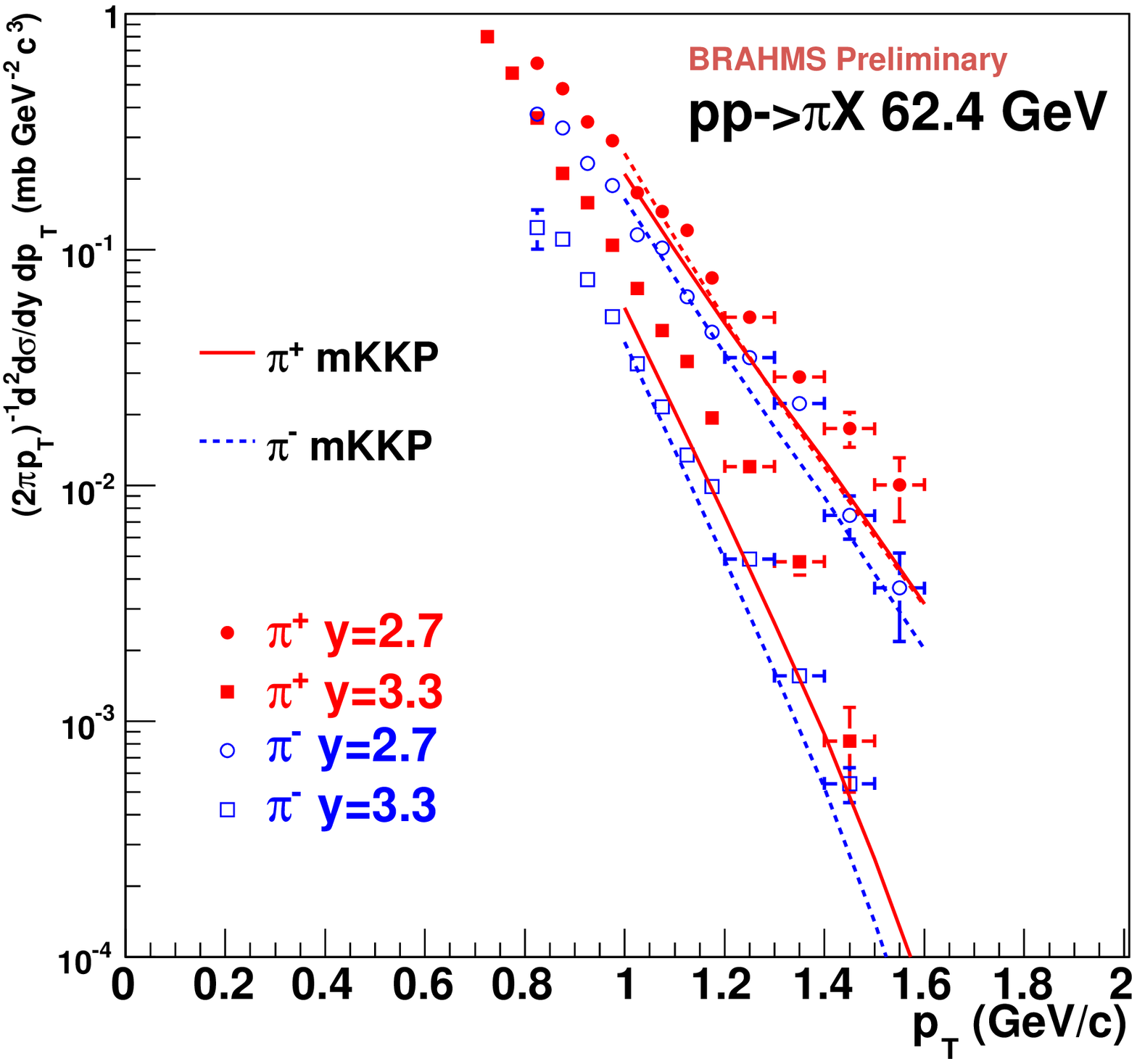}
\end{center}
\caption{Invariant cross section for midrapidity ($|\eta|<0.35$) neutral pion (left) and forward charged pion (right) production at $\sqrt{s}=62.4$~GeV, compared to perturbative QCD calculations.}
\label{fig:xsect62}
\end{figure}

Polarization-averaged cross sections for pion production in proton-proton collisions at 200~GeV at both midrapidity as well as the forward region have been measured at RHIC and found to be well described by next-to-leading-order (NLO) pQCD  \cite{Adare:2007dg,Adams:2006uz,Arsene:2007jd}.  Pion cross section measurements have also been performed at RHIC for $\sqrt{s}=62.4$~GeV at mid- and forward rapidities \cite{Adare:2008cx,Videbaek:2008ht} and are shown in Fig.~\ref{fig:xsect62} compared to pQCD calculations.  On the left one can see the cross section for midrapidity neutral pions as a function of transverse momentum ($p_T$), compared to NLO as well as next-to-leading-log (NLL) pQCD calculations performed using $p_T$ as the choice of renormalization and factorization scale.  The bottom panels indicate the differences obtained in the calculations if scale choices are instead $p_T/2$ or $2p_T$.  On the right side of Fig.~\ref{fig:xsect62} are shown the cross sections versus $p_T$ for positive and negative pions measured at rapidities of 2.7 and 3.3, compared to NLO pQCD calculations.  While the description of the data is not as successful as at midrapidity, the agreement is still reasonably good.

\section{The PHENIX and BRAHMS experiments at RHIC}
\subsection{PHENIX}
PHENIX is one of two large, ongoing experiments studying proton spin structure at RHIC.  The PHENIX detector \cite{Adcox:2003zm} consists of two multipurpose spectrometer arms at midrapidity ($|\eta|<0.35$) covering a total of $\pi$ in azimuth and two larger-rapidity spectrometer arms covering $1.2 < |eta| < 2.4$ and $\Delta \phi = 2\pi$ dedicated to the measurement of muons.  The midrapidity spectrometers are instrumented to track and identify charged particles as well as detect electromagnetic probes.  In 2006-07 two large-rapidity electromagnetic calorimeters covering $3.1 < |\eta| < 3.7$, the muon piston calorimeters (MPC), were added.  The PHENIX experiment was designed to detect rare probes, with fast data acquisition (greater than 5~kHz in proton-proton running) and sophisticated triggering capabilities.

The stable direction of the spin of the proton beams as they circulate in the RHIC ring is vertical.  Spin rotator magnets immediately outside the PHENIX interaction region allow for the choice between transverse and longitudinal polarization of the colliding beams, independent of other RHIC experiments.

\subsection{BRAHMS}

The BRAHMS experiment, which finished taking data in 2006, consists of two movable spectrometer arms for the measurement of identified charged hadrons over a wide range of rapidity and transverse momentum \cite{Adamczyk:2003sq}.  The Forward Spectrometer (FS), which can be rotated from 2.3. to 15. degrees relative to the beam line, consists of 4 dipole magnets with a bending power of up to 9.2 Tm.  There are five tracking stations, and a segmented time-of-flight wall as well as a ring-imaging Cherenkov detector are used for particle identification.  The Mid-Rapidity Spectrometer (MRS), which can be rotated from 34. to 90. degrees relative to the beam line, is a single-dipole-magnet spectrometer with a solid angle of approximately 5~msr and a magnetic bending power up to 1.2 Tm.  The MRS contains two time projection chambers followed by two highly segmented scintillator time-of-flight walls.

With no spin rotator magnets outside the BRAHMS interaction region, all proton-proton collisions at BRAHMS are transversely polarized in the vertical direction.

\section{Results}
A number of results are now available from transversely polarized data taken by the BRAHMS and PHENIX experiments at center-of-mass energies of 200 and 62.4~GeV.  The transverse single-spin asymmetries discussed below are all left-right asymmetries, which can be calculated by
\begin{equation} \nonumber
A_N^{\rm Left} = \frac{1}{P} \frac{N^\uparrow - RN^\downarrow}{N^\uparrow + RN^\downarrow}
\end{equation}
where $A_N^{\rm Left}$ indicates the asymmetry calculated to the left of the polarized beam, $P$ is the beam polarization, $N^\uparrow$ ($N^\downarrow$) is the particle yield from bunches polarized up (down), and $R = \frac{L^\uparrow}{L^\downarrow}$ is the relative luminosity between up- and down-polarized bunches.  Both beams at RHIC are polarized; in the calculation of single-spin asymmetries, the polarization of one beam is considered while averaging over the polarization states of the other.

\begin{figure}
\begin{center}
\includegraphics[width=0.65\textwidth,height=0.3\textheight]{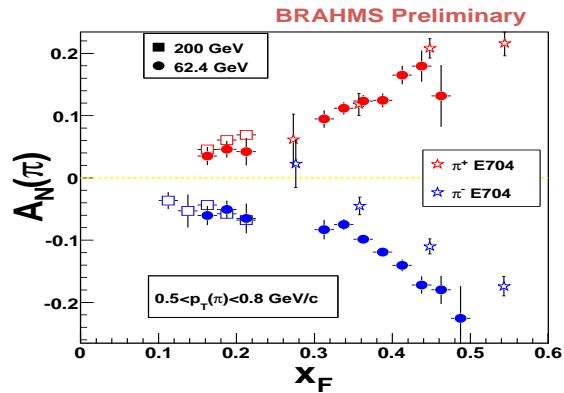}
\vspace{7 mm}
\end{center}
\vspace{-7 mm}
\caption{Charged pion asymmetries measured at 200 and 62.4~GeV by the BRAHMS experiment and at 19.4~GeV by the E704 experiment, shown for overlapping kinematic ranges (see text).}
\label{fig:threeEnergies}
\end{figure}

In the early 1990's large transverse single-spin asymmetries in forward pion production were observed by the E704 experiment at Fermilab at a center-of-mass energy of 19.4~GeV \cite{Adams:1991cs,Adams:1991rw}.  The asymmetries of the charged pion species were found to be of similar magnitude with opposite sign, while the neutral pions exhibited a positive asymmetry, smaller in magnitude.  The magnitude of the asymmetries increased with increasing Feynman-$x$ ($x_F = 2p_L/\sqrt{s}$).  Transverse SSA's in pion production at RHIC were initially measured by the STAR experiment at $\sqrt{s}=200$~GeV \cite{Adams:2003fx}.  The asymmetry in forward neutral pion production was found to be positive and increasing with $x_F$, as at E704.  Subsequent measurements of forward charged pions by BRAHMS at 200~GeV revealed the same pattern as observed at 19.4~GeV: positive and negative pions exhibit near mirror symmetry in their behavior versus $x_F$, with a larger magnitude than the neutral pions.  In 2006, data at the intermediate energy of $\sqrt{s}=62.4$~GeV was taken at RHIC, and measurements of charged pions by BRAHMS \cite{Arsene:2008mi} and neutral pions by PHENIX \cite{Chiu:2007zy} again demonstrated the same pattern.  In Fig.~\ref{fig:threeEnergies}, the asymmetries for charged pions at all three energies are shown in overlapping kinematic ranges. The data at 62.4 and 200~GeV are for a transverse momentum ($p_T$) selection of 0.5-0.8 GeV/$c$; the 19.4~GeV data is for $p_T > 0.7$~GeV/$c$.  The positive pion points overlap at all three energies, suggesting some kind of scaling behavior.  The negative pion asymmetry at the lowest energy, however, appears to have a somewhat smaller magnitude than at the two higher energies.  In any case, the similarities among the three energies are striking and suggest a common origin.

\begin{figure}
\begin{center}
\includegraphics[width=1.0\textwidth]{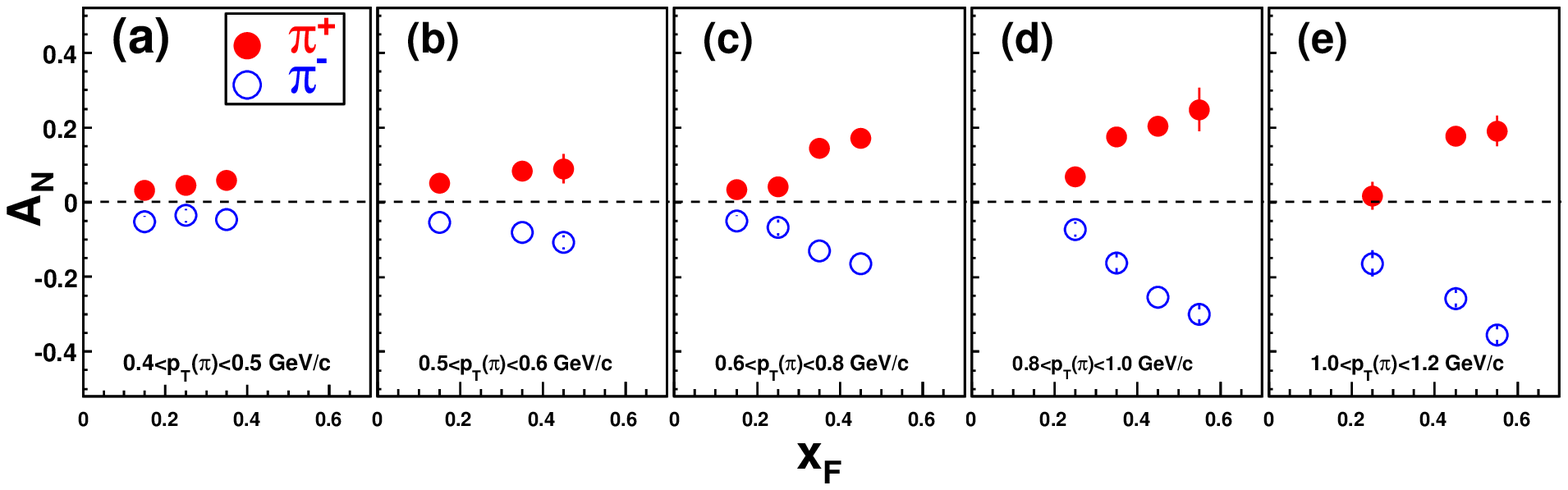}
\includegraphics[width=0.60\textwidth,height=0.25\textheight]{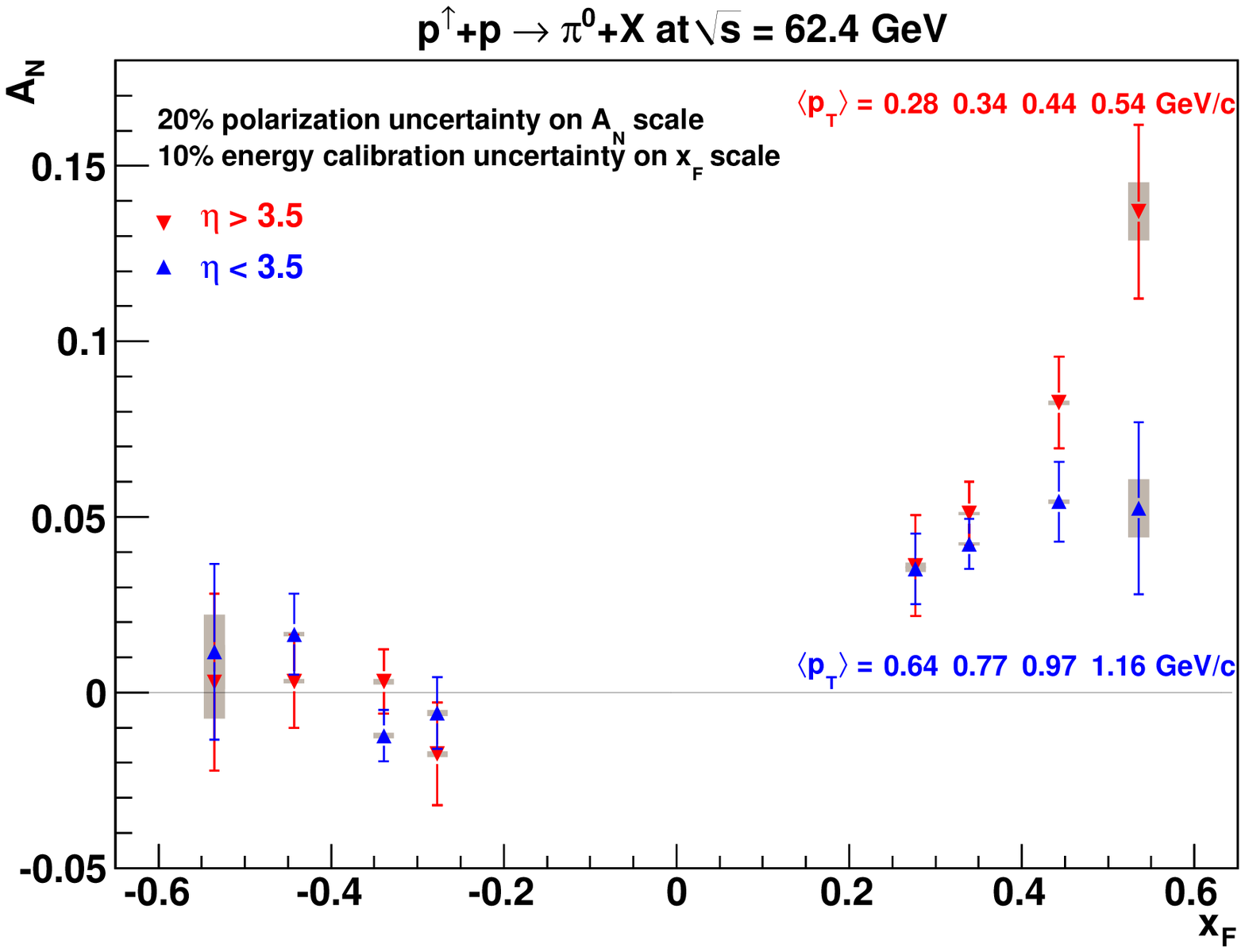}
\end{center}
\vspace{-5 mm}
\caption{Transverse single-spin asymmetry of charged and neutral pions for $\sqrt{s}=62.4$~GeV in bins of transverse momentum, measured by BRAHMS \cite{Arsene:2008mi} (top), and for two different pseudorapidity ranges, measured by PHENIX \cite{Chiu:2007zy} (bottom).}
\label{fig:mpc}
\end{figure}

The behavior of the pion asymmetries has also been studied in bins of $p_T$ and rapidity, as shown for example in Fig.~\ref{fig:mpc} for charged and neutral pions at 62.4~GeV.  The magnitude of the charged pion asymmetries appears to increase with increasing transverse momentum up to a $p_T$ of approximately 1.0~GeV/$c$; however, the neutral pion data imply an opposite trend.  Additional high-statistics data allowing detailed study of the asymmetries as a function of various kinematic variables will be important to clarify the trends currently observed as well as to pin down theoretical models which can correctly describe the behavior of the asymmetries in terms of multiple kinematic variables.  At present, several effects are believed to be potential contributors to the observed asymmetries, with existing descriptions based on twist-three, Sivers, and Collins mechanisms \cite{Qiu:1998ia,Sivers:1989cc,Collins:1992kk}.  Contrary to a prior claim \cite{Anselmino:2004ky}, the magnitude of the pion asymmetries observed at at RHIC at 200~GeV can in fact be explained by the Collins mechanism \cite{Yuan:2008tv,Murgia:2008tp}.

\begin{figure}
\psfig{file=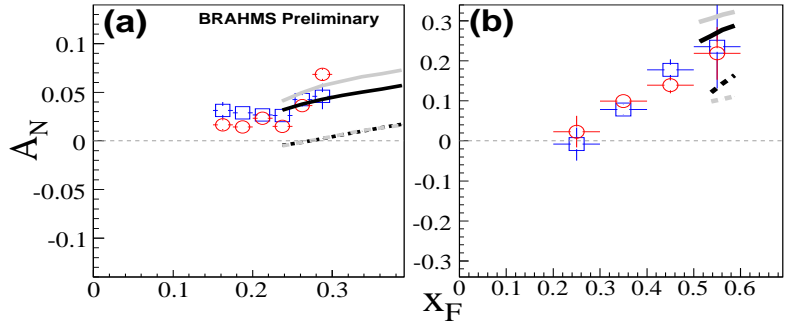,width=0.92\textwidth,height=0.25\textheight}
\caption{Transverse SSA in kaon production at 200~GeV (left) and 62.4~GeV \cite{Arsene:2008mi} (right), measured by BRAHMS.  Positive (negative) kaons are indicated by open circles (squares). Solid (dashed) curves are for positive (negative) kaon initial-state twist-three calculations with (black) and without (grey) sea-quark contributions.}
\label{fig:kaons}
\end{figure}

The BRAHMS experiment has also measured transverse SSA's in kaon production at 62.4 and 200~GeV, and as the pions, they exhibit similar behavior at the two energies.  As can be seen in Fig.~ \ref{fig:kaons}, the asymmetry of both charges is positive and of a similar magnitude.  As shown in the figure, current theoretical calculations underpredict the negative kaon asymmetries.

\begin{figure}
\begin{center}
\includegraphics[width=0.42\textwidth,height=0.25\textheight]{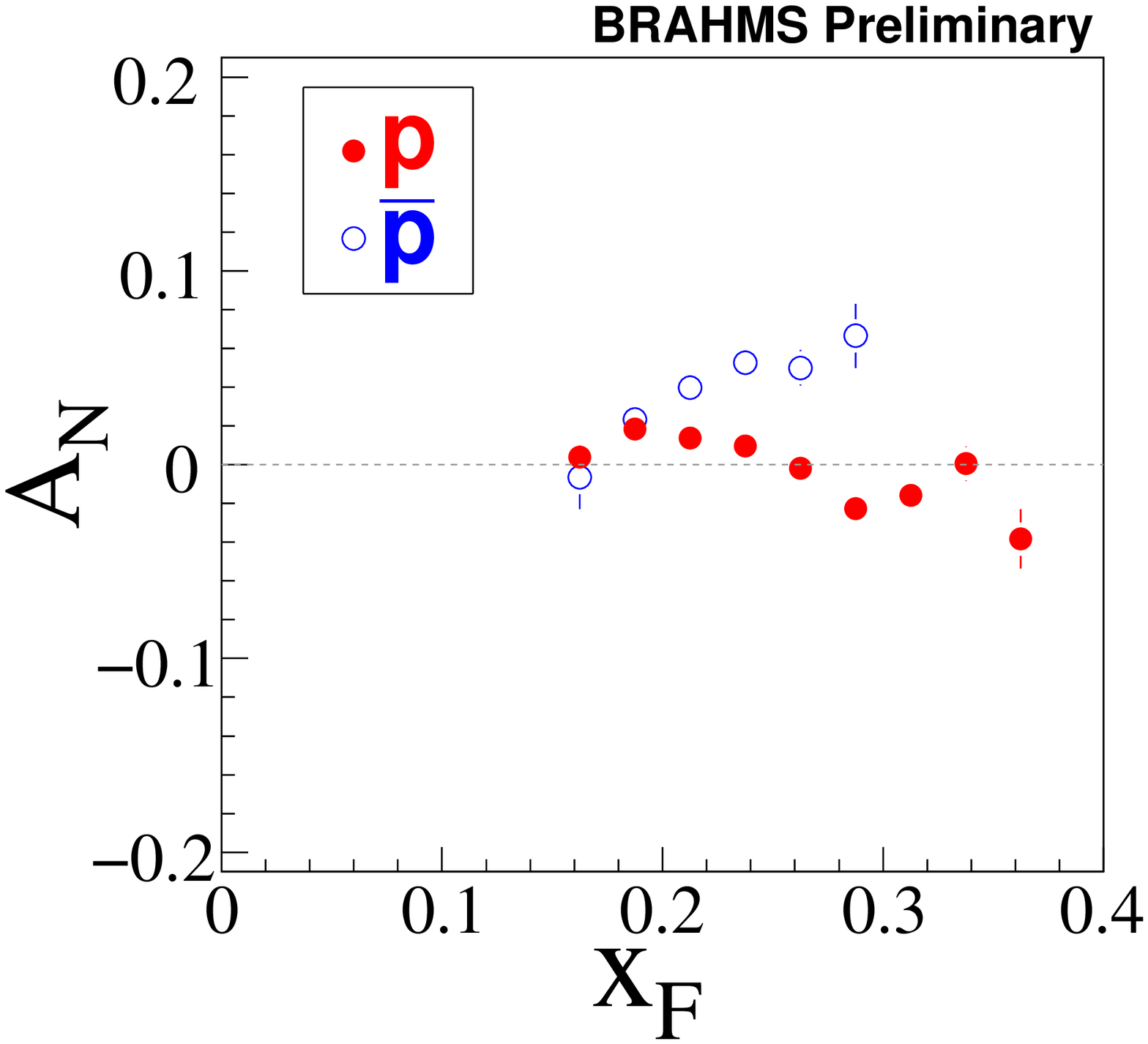}
\hspace{0.03\textwidth}
\includegraphics[width=0.42\textwidth,height=0.25\textheight]{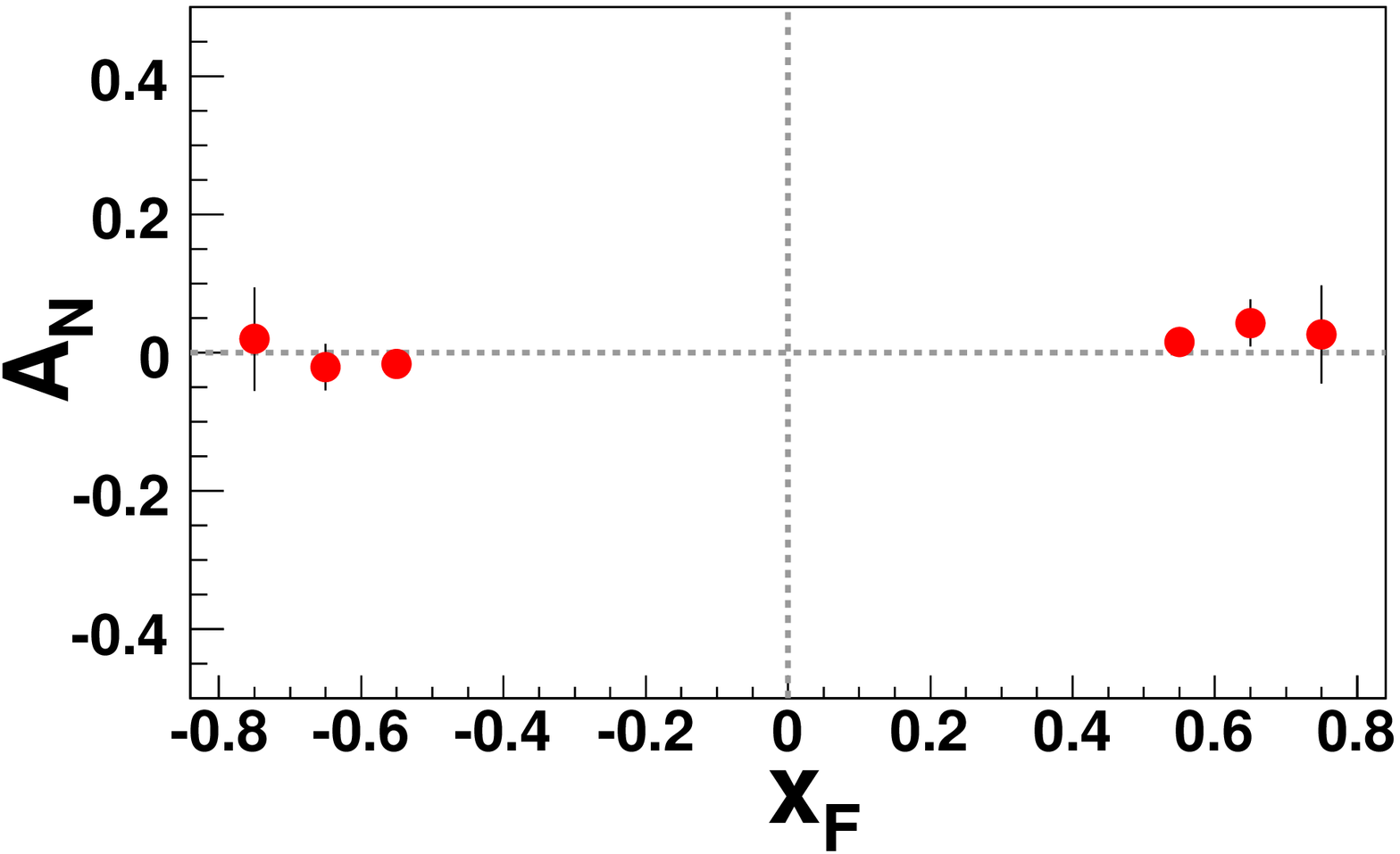}
\end{center}
\vspace{-5 mm}
\caption{Transverse SSA in proton and antiproton production at 200~GeV (left) and 62.4~GeV \cite{Arsene:2008mi} (right), measured by BRAHMS.  The 62.4~GeV points are for protons only.}
\label{fig:protons}
\end{figure}

In addition to pions and kaons, BRAHMS has also measured proton asymmetries, as shown in Fig.~\ref{fig:protons}.  While the proton asymmetry is consistent with zero for both energies, the antiprotons at 200~GeV demonstrate a clear positive asymmetry.  Unfortunately the data for a measurement of antiprotons at 62.4~GeV are not available.  These surprising results remain completely unexplained.

In contrast to the forward pion and kaon asymmetries, a measurement of neutral pions and charged hadrons at midrapidity for $1<p_T<5$~GeV/$c$ was performed by PHENIX and found to be consistent with zero within a few percent \cite{Adler:2005in}.  Improvement of a factor of approximately 200 compared to the results published is anticipated in measurements at PHENIX of the transverse SSA of midrapidity neutral pions based on data taken in 2006 and 2008.  The published result is sensitive mainly to gluon-gluon scattering and has been used to constrain the gluon Sivers function \cite{Anselmino:2006yq}.  The more recent data are expected to provide at low $p_T$ an even stronger constraint on the gluon Sivers function and at higher $p_T$ increased sensitivity to quark effects.

\begin{figure}
\begin{center}
\includegraphics[width=0.42\textwidth,height=0.25\textheight]{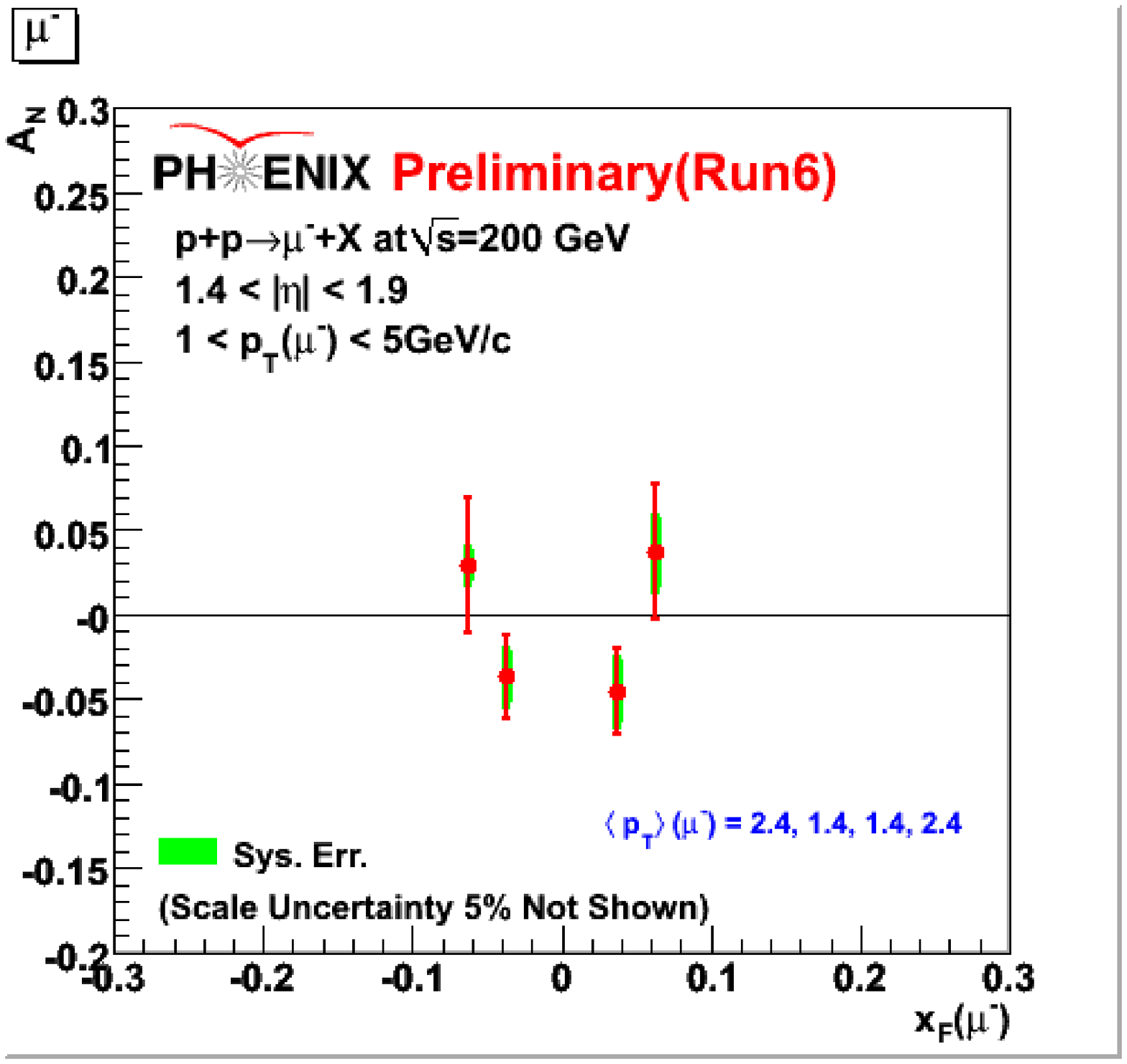}
\hspace{0.03\textwidth}
\includegraphics[width=0.42\textwidth,height=0.25\textheight]{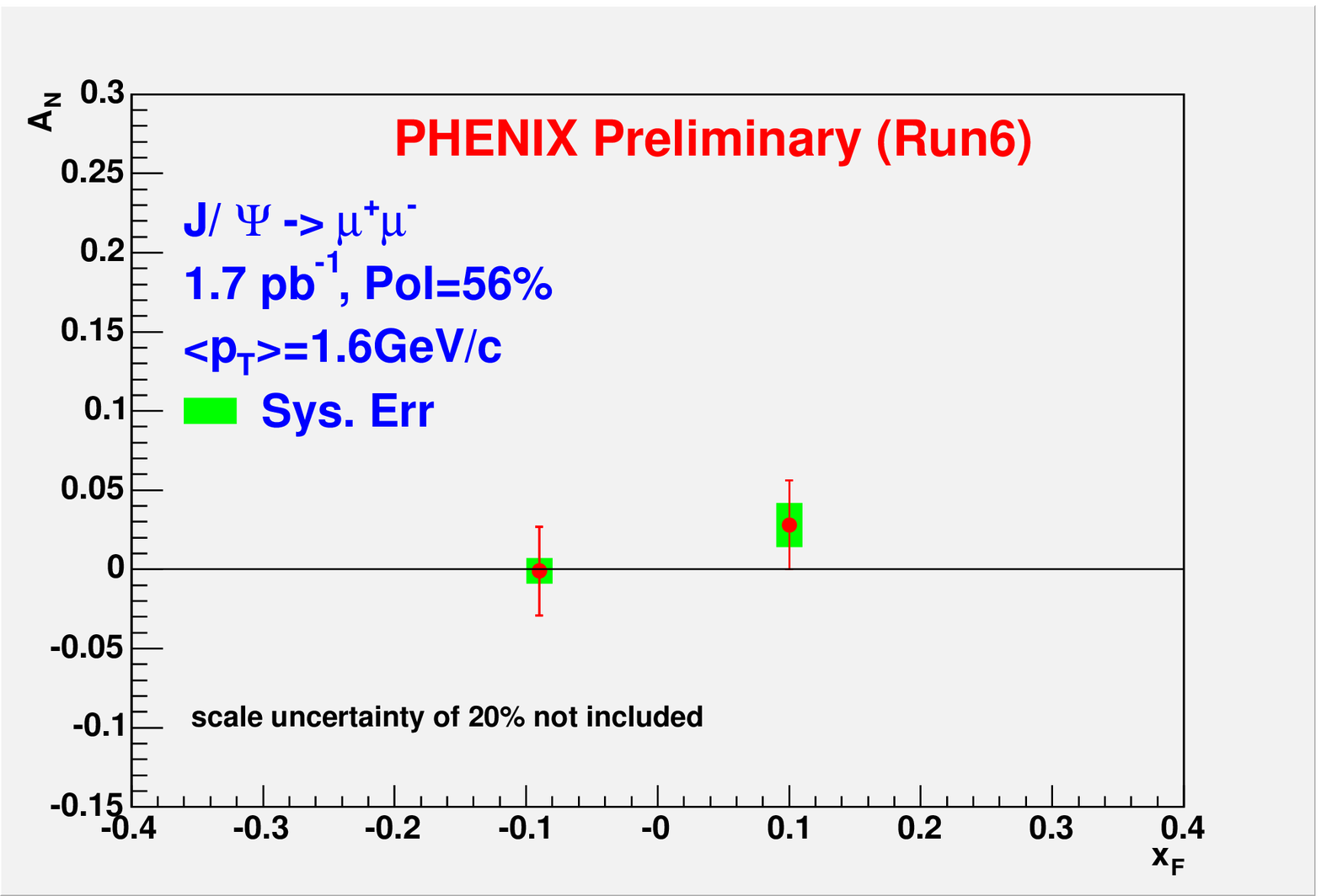}
\end{center}
\vspace{-2 mm}
\caption{Transverse SSA measured by PHENIX for $\sqrt{s}=200$~GeV in single muons from heavy flavor decays (left) and J/$\Psi$ production (right).}
\label{fig:muons}
\end{figure}

It was proposed several years ago that the transverse SSA in $D$ meson production would be an excellent channel to probe the gluon Sivers function \cite{Anselmino:2004nk}.  A first measurement has now been made of the transverse SSA in the production of single muons from heavy flavor decays, shown versus $x_F$ in Fig.~\ref{fig:muons}.  Results are also available as a function of transverse momentum.  The significant systematic uncertainties indicated in the figure are due mainly to the estimation of the background contribution and should be improved in the future by the planned forward silicon detector upgrade to PHENIX.  A direct comparison between the calculations of Anselmino \emph{et al.} \cite{Anselmino:2004nk}, done for $D$ mesons, and the present results, which are for single muons resulting from the decay of both charm and bottom mesons, is not currently possible. However, it is worth noting that the magnitude of the observed muon asymmetries at positive $x_F$ is significantly smaller than the calculation of 10-15\% asymmetries in a similar $x_F$ and $p_T$ range for $D$ mesons.  We thus hope that with some further work in translating between $D$ mesons and their decay muons it will be possible to use these data to place a stronger constraint on the gluon Sivers function than presently available.

Looking further at what we can learn from transverse single-spin asymmetries in heavy flavor measurements, recent work indicates that the transverse SSA in J/$\Psi$ production may be sensitive to the J/$\Psi$ production mechanism \cite{Yuan:2008vn}.  A non-zero asymmetry is predicted in $p+p$ collisions in the color-singlet model; a zero asymmetry in the color-octet model.  This is in contrast to J/$\Psi$ production in semi-inclusive DIS, in which a non-zero transverse single-spin asymmetry is predicted in the color-octet model but zero asymmetry in the color-singlet model.  Preliminary results on the J/$\Psi$ transverse SSA in $p+p$ collisions can be seen in Fig.~\ref{fig:muons}.  Early estimates suggest that an asymmetry of the order of a few percent at positive $x_F$, consistent with the data, may be possible due to \emph{indirect} J/$\Psi$ production coming from feed-down from the $\chi_c$, assuming a gluon Sivers function of approximately $0.5x(1-x)xG(x)$ \cite{Yuan:2008pr}.  However, more detailed theoretical work will be needed to clarify what can be learned from the current experimental result; improvement in the experimental measurement is expected once analysis of the 2008 transverse data set from PHENIX has been completed.

\begin{figure}
\begin{center}
\psfig{file=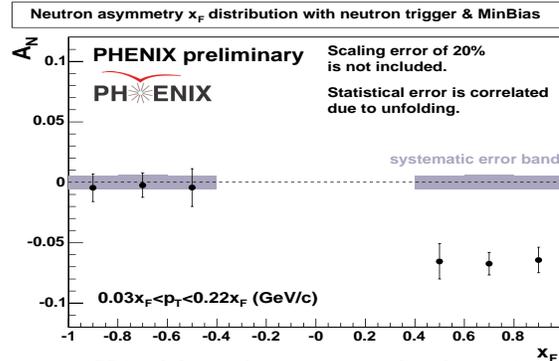,width=0.72\textwidth,height=0.27\textheight}
\end{center}
\vspace{-6 mm}
\caption{Transverse SSA of forward neutrons produced in coincidence with forward charged particles for $\sqrt(s)=200$~GeV, measured by PHENIX.}
\label{fig:neutrons}
\end{figure}

A large, negative asymmetry in the production of very forward neutrons at $\sqrt{s}=200$~GeV has been observed by the PHENIX experiment \cite{Togawa:2007zza}.  This asymmetry has been found to be enhanced by requiring coincidence with forward charged particles in the pseudorapidity range of $3.0 < |\eta| < 3.9$ (Fig.~\ref{fig:neutrons}).  As can be seen in the figure, the forward asymmetry exhibits no dependence on $x_F$, while the asymmetry in the backward direction is consistent with zero.  A large negative transverse SSA in the production of forward neutrons at $\sqrt{s} = 200$~GeV has also been observed by a small experiment located at RHIC Interaction Point 12 \cite{Bazilevsky:2006vd}, and similar asymmetries have been observed by PHENIX for center-of-mass energies of 62.4 as well as 410~GeV.  Unlike other results presented above, the kinematics of the neutron measurement are not in a regime described by perturbative processes.  The large measured asymmetries suggest interference between a spin-flip amplitude, such as one-pion exchange, which dominates lower-energy neutron production, and non-flip amplitudes, but a clear understanding of the large asymmetries observed has yet to be developed.

\section{Conclusions}
A wealth of transverse single-spin asymmetry measurements is now available from the PHENIX and BRAHMS experiments at RHIC, the majority of which are in a kinematic region in which the polarization-averaged cross sections are well described by pQCD.  For inclusive pion measurements, enough statistics are available to allow first studies of the asymmetry dependence on $p_T$ and rapidity in addition to $x_F$.  The behavior of the SSA's as a function of various kinematic variables is essential to constrain phenomenological models of the underlying physics generating the asymmetries.  With first extractions of the Collins fragmentation function as well as the transversity and Sivers distributions now available thanks to measurements in $e^+e^-$ and SIDIS, an increasing amount can be learned from proton-proton collision data.  The long future of the RHIC spin program promises a myriad of further transverse-spin results, and hadronic data will play a crucial role in our understanding of transversity and transverse-momentum-dependent distributions as the field comes to maturity.

\bibliographystyle{ws-procs9x6}
\bibliography{AidalaTransversity2008}

\end{document}